\documentstyle[psbox]{WORKSHOP}
\begin{document}

\title{
Origin of the 6.4-keV line of the Galactic Ridge X-ray Emission \\ 
{\large\sf  -- First Report --} 
}

\author{
T.G.Tsuru$^1$, 
H.Uchiyama$^2$, 
K.K.Nobukawa$^1$, 
M.Nobukawa$^1$, \\
S.Nakashima$^3$, 
K.Koyama$^1$, 
K.Torii$^4$, 
Y.Fukui$^4$, 
\\[12pt]  
%
$^1$  Department of Physics, Kyoto University, Kitashirakawa, Sakyo, Kyoto, JAPAN 606-8502 \\
$^2$  Faculty of Education, Shizuoka University, 836 Ohya, Shizuoka, JAPAN 422-8529  \\
$^3$  ISAS, JAXA,  3-1-1 Yoshinodai, Chuo-ku, Sagamihara, Kanagawa, JAPAN 252-5210\\
$^4$  Department of Astrophysics, Nagoya University, Furo-cho, Chikusa-ku, Nagoya, JAPAN 464-8602\\
%
{\it E-mail(TGT): tsuru@cr.scphys.kyoto-u.ac.jp} 
}

\abst{
We report the first results from high-statistics observation of the 6.4-keV line in the region of $l= +1.5^\circ$ to $+3.5^\circ$ (hereafter referred to as GC East), with the goal to uncover the origin of the Galactic ridge X-ray emission (GRXE). 
By comparing this data with that from the previous observations in the region $l=-1.5^\circ$ to $-3.5^\circ$ (hereafter referred to as GC West), we discovered that the 6.4-keV line is asymmetrically distributed with respect to the Galactic center, whereas the 6.7-keV line is symmetrically distributed. 
The distribution of the 6.4-keV line follows that of $^{13}$CO and its flux is proportional to the column density of the molecular gas.  
This correlation agrees with that seen between the 6.4-keV line and the cold interstellar medium (ISM) (H$_{\rm I}$ $+$ H$_2$) in the region $|l|>4^\circ$. 
This result suggests that the 6.4-keV emission is diffuse fluorescence from the cold ISM not only in GC East and West but also in the entire Galactic plane. 
This observational result suggests that the surface brightness of the 6.4-keV line is proportional to the column density of the cold ISM in the entire Galactic plane. 
For the ionizing particles, we consider X-rays and low energy cosmic-ray protons and electrons . 
}

\kword{Galaxy: center --- Interstellar medium --- X-ray spectra}

\maketitle
\thispagestyle{empty}

\section{Introduction: Neutral iron K line in the Galactic ridge}
Suzaku successfully resolved the structure of the highly ionized iron K emission lines from the Galactic center region and determined the plasma temperature, not from the shape of the continuum spectrum, but from the ratio of the fluxes of the 6.7-keV line (He-like ions) to 6.9-keV line (H-like ions), thanks to its large collecting area and precise spectroscopic performance. 
Koyama et~al. (2007) and Yamauchi et~al. (2009) found that the plasma temperature ($kT \sim 5.5 {\rm keV}$) of the Galactic ridge X-ray emission (GRXE, $|l| > 1.5^\circ$) is significantly lower than the plasma temperature ($kT \sim 7 {\rm keV}$) of the Galactic center diffuse X-ray emission (GCDX, $|l| < 1.5^\circ$). 
Thus, these thermal emissions should have different origins. 

Figure~1 shows the equivalent width map of the 6.4-keV line (neutral iron) in and near the Galactic center region, in which the prominent features of Sagittarius (Sgr) A, B, C, D, and E are seen. Sgr~D and E are new discoveries by Suzaku (Ryu 2013). 
The idea of X-ray reflection nebula irradiated by Sgr~A* is now widely accepted. 
No region with such high equivalent width has been discovered except the Galactic center region. 
However, Suzaku observed uniform 6.4-keV line emission with nearly equal equivalent width that extends toward the Galactic ridge, where it also discovered the 6.4-keV line from GRXE (Ebisawa et al. 2008; Yamauchi et al. 2009).  
In addition, we found a hint that the 6.4-keV line emission from the GRXE to the east-side of the Galactic center region is stronger than that to the west-side (Maeda 1998; Uchiyama et al. 2011). 
Here, we present the first report of the high-statistics observation of the 6.4-keV line in the region of $l = +1.5^\circ \sim +3.5^\circ$ (hereafter referred to as GC East), which was done in the Suzaku AO-7 key project. 

\begin{figure*}[t]
\centering
\psbox[xsize=12cm]
{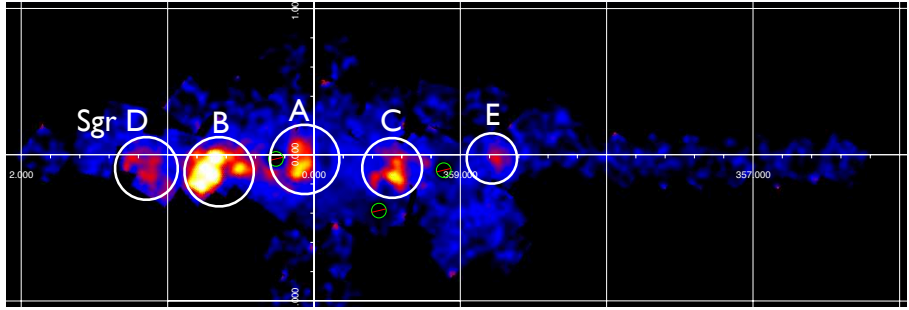}
\caption{Equivalent width map of the 6.4-keV line in and near the Galactic center region obtained with Suzaku.}
\end{figure*}

\section{Observation}

Our AO-7 project observation consists of 10 pointings of 100~ksec exposure each at the Galactic longitude from $l= +1.5^\circ$ to $+4.0^\circ$. 
We had finished 7 pointings by February 2014 and were planned to do remaining three pointings during the spring 2014. 
Figure~2 shows mosaic images including the previous observation in the eight X-ray energy bands from 0.5 to 8~keV. 
Artificial structures are apparent in the energy bands from 1 to 7~keV. 
These structures are due to stray X-ray light from GX~$3+1$, which is a very bright low-mass X-ray binary located at $ (l,b)=(+2.294^\circ, +0.7937^\circ)$ (Mori et al. 2005). Thus, careful analyses are necessary. 

\begin{figure}[t]
\centering
\psbox[xsize=8cm]
{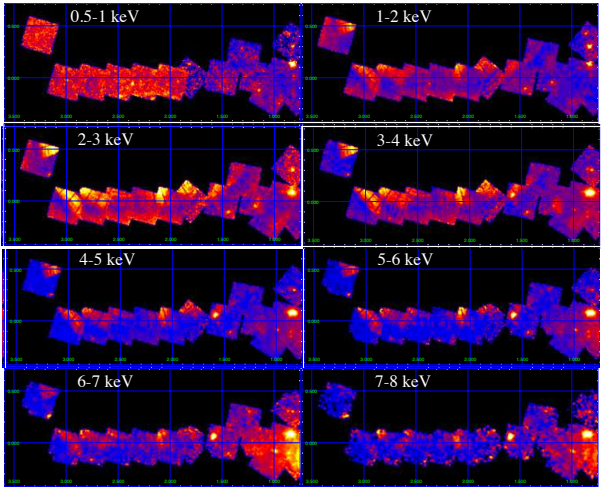}
\caption{X-ray mosaic mages in the eight energy bands from $0.5$ to $8$~keV at the Galactic longitude of $l=1^\circ$ to $3.5^\circ$. }
\end{figure}

\subsection{Stray Light from GX~$3+1$}

Figure~3 shows an typical image. 
The image is divided into three regions according to the type of stray light (Mori et al. 2005; Suzaku TD). 
Region (a) is called the ``secondary-reflection region'', where the X-rays from the stray source are reflected only once by the secondary reflector. 
This region is the brightest of the three and a comb-like structure is clearly seen. 
Region (b) is the ``backside-reflection region'', where the X-rays from the stray source are scattered at the backside of the primary reflector and are followed by the normal double reflection.
The iron K line band is free from stray light because stray light is only bright below 1.5~keV. 
Region (c) is referred to as the ``quadrant-boundary region'' and is the shadow of the structure of a quadrant of the XRT of the Suzaku satellite. 
This region is free from stray light.

\begin{figure}[t]
\centering
\psbox[xsize=6cm]
{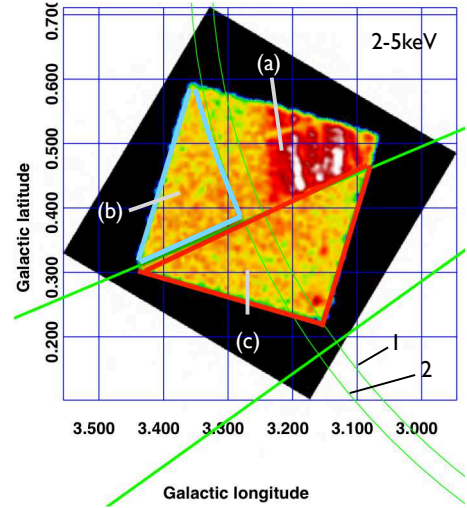}
\caption{
Example of X-ray images obtained in the Suzaku key project observation is reported here. 
The regions marked with (a), (b), and (c) are the secondary-reflection, backside-reflection, and quadrant-boundary regions, respectively  (Mori 2005; Suzaku TD). 
The region between the two green lines is the shadow of the structure of the quadrant of the XRT of the Suzaku satellite. 
The both arcs of 1 and 2 share a common center at GX~3+1. 
Arc 1 passes the field of view center of this observation and arc 2 has a radius 2 arcmin larger than arc 1.
Stray light comes only from the backside-reflection outside of arc 2.}
\end{figure}

\subsection{The fluxes of the 6.4- and 6.7-keV lines}

The stray light in the secondary-reflection region is too bright to obtain the continuum component of the GRXE. 
However, observation by XMM-Newton revealed no narrow iron K line in the spectrum of GX~$3+1$ (Piraino et al. 2012). 
Thus, we were able to use the entire field of view (FOV) to obtain the flux of iron K lines. 
For confirmation, we compare the iron flux obtained from the entire FOV with the iron flux in the quadrant-boundary region plus in the backside-reflection region, where the iron K line band is free from stray light. 
Figure~4 shows the flux of 6.7- and 6.4-keV lines as a function of the Galactic longitude. 
No significant difference between the two regions is observed. 
For good statistics, we adopt the iron K line flux from the entire FOV in the following analyses. 

\begin{figure}[t]
\centering
\psbox[xsize=8cm]
{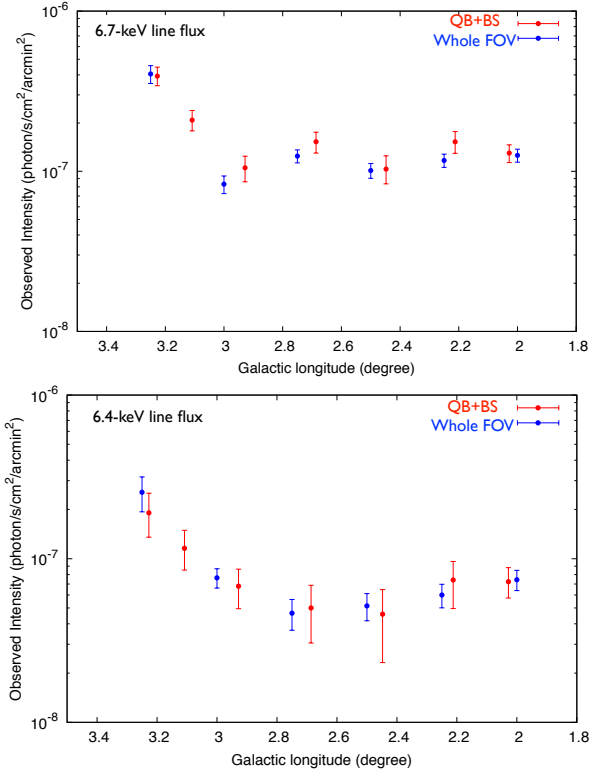}
\caption{
The surface brightness of the 6.7-keV (upper panel) and 6.4-keV (lower panel) lines as a function of the Galactic longitude. 
Blue and red points represent the data obtained in the entire FOV region and the region of quadrant-boundary region plus the backside-reflection region, respectively.}
\end{figure}

\section{Results and Discussion}

\subsection{Asymmetrical distribution of the 6.4-keV line}
Figure~5 shows the flux of the iron K emission lines as a function of Galactic longitude. 
This figure includes the data collected during the previous observation. 
Although the 6.7-keV line is symmetrically distributed with respect to the Galactic center, the 6.4-keV line is asymmetrically distributed. 
The flux of the 6.4-keV line in GC East is twice that in the region $l=-1.5^\circ$ to $3.5^\circ$ (hereafter referred to as GC West). 
If the 6.4-keV line has the same origin as the 6.7-keV line, the two lines should have same symmetrical distribution. 
Thus, the observational results suggest that the origin of the 6.4-keV line is different from that of the 6.7-keV line. 
Thus, GRXE consists comprises at least two components; one emitting the 6.4 keV line and the other emitting the 6.7 keV line.

\begin{figure*}[t]
\centering
\psbox[xsize=12cm]
{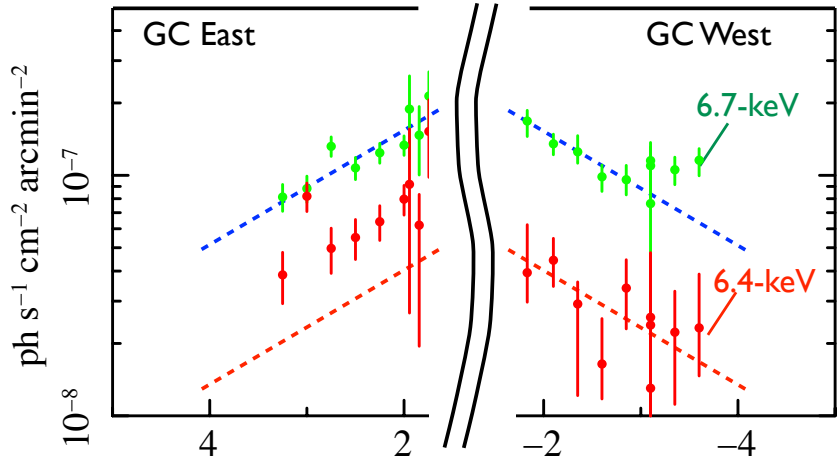}
\caption{
Surface brightness of 6.4-keV line (red) and 6.7-keV line (green) as a function of the Galactic longitude. 
Red and blue dashed lines are to guide the eye and show the symmetry with the respect to the Galactic center.}
\end{figure*}

\subsection{Distribution of 6.4-keV line and molecular clouds}
The 6.4-keV line is due to fluorescence from molecular clouds in the Galactic center region. 
Thus, using the data obtained with the NANTEN telescope (Takeuchi et al. 2010), 
we compare the distribution of the 6.4-keV line in GC East and West with that of the molecular clouds. 
Figure~6 shows the surface brightnesses of $^{13}$CO and the 6.4-keV line as a function of the Galactic longitude. 
The surface brightness of $^{13}$CO is the average from $b=-0.126^\circ$ to$+0.034^\circ$, whereas the surface brightness of the 6.4-keV line is the average over the FOV. 
The 6.4-keV line follows the $^{13}$CO surface brightness very well, which suggests that the 6.4-keV line is related to the molecular clouds. 
The structure at $l=3.2^\circ$ is the molecular cloud complex called ``Clump 2'', which extends from $(l, b) = (+3.2^\circ, 0.0^\circ)$ to $(+3.2^\circ ,+0.5^\circ)$ (Bania et al. 1977). 
In this region, the 6.4-keV line also follows the $^{13}$CO surface brightness. 

\begin{figure*}[t]
\centering
\psbox[xsize=12cm]
{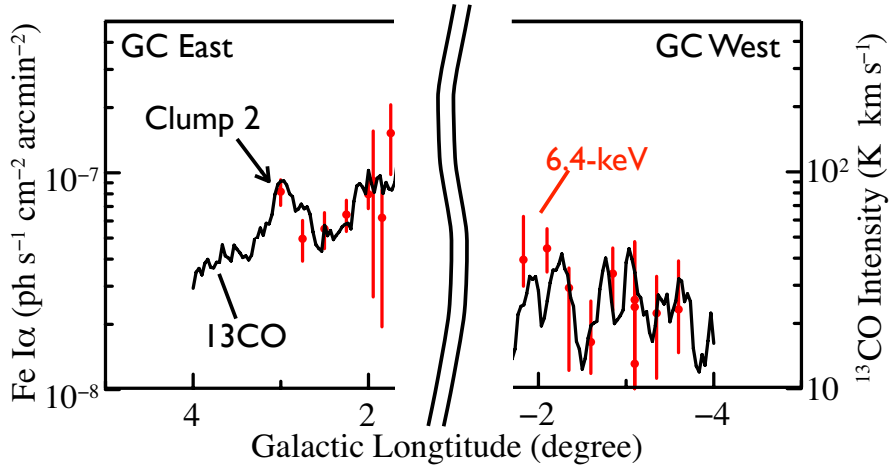}
\caption{The surface brightness of the 6.4-keV line (red) and intensity of $^{13}$CO (black) as a function of galactic longitude. }
\end{figure*}

\subsection{Correlation between the 6.4-keV line and cold ISM in Galactic plane}
Figure~7 shows the surface brightness of the 6.4-keV line as a function of cold ISM column density in GC East and West. 
For this figure, we applied the conversion factor of 
$N_{\rm H} / W(^{13}{\rm CO}) = 1\times 10^{23}{\rm cm^{-2}} / 57.1\ {\rm K\ km\ sec^{-1}}$. 
Note that the cold ISM is dominated by molecular clouds in GC East and West. 
The figure shows that the flux of the 6.4-keV line is proportional to the cold ISM column density. 
This result suggests that the origin of the 6.4-keV line is diffuse and could be fluorescence from the cold ISM. 

We now discuss the entire Galactic plane beyond GC East and West. 
Uchiyama et~al. (2014) recently found good correlation between the surface brightness of the 6.4-keV line and the cold ISM column density consisting of H$_{\rm I}$ (Kalberla et~al. 2005) and H$_2$ calculated from the CO map by Dame et~al. (2001) in the region $|l|>4^\circ$ in the Galactic longitude, where the contribution from H$_{\rm I}$ gas to the cold ISM is significant, and molecular gas (H$_2$) as well. 
We find that the correlations are very consistent with each other (Figure~7), which suggests that the 6.4-keV emission from the entire Galactic plane has the same diffuse origin as in GC East and West. 

\begin{figure}[t]
\centering
\psbox[xsize=8cm]
{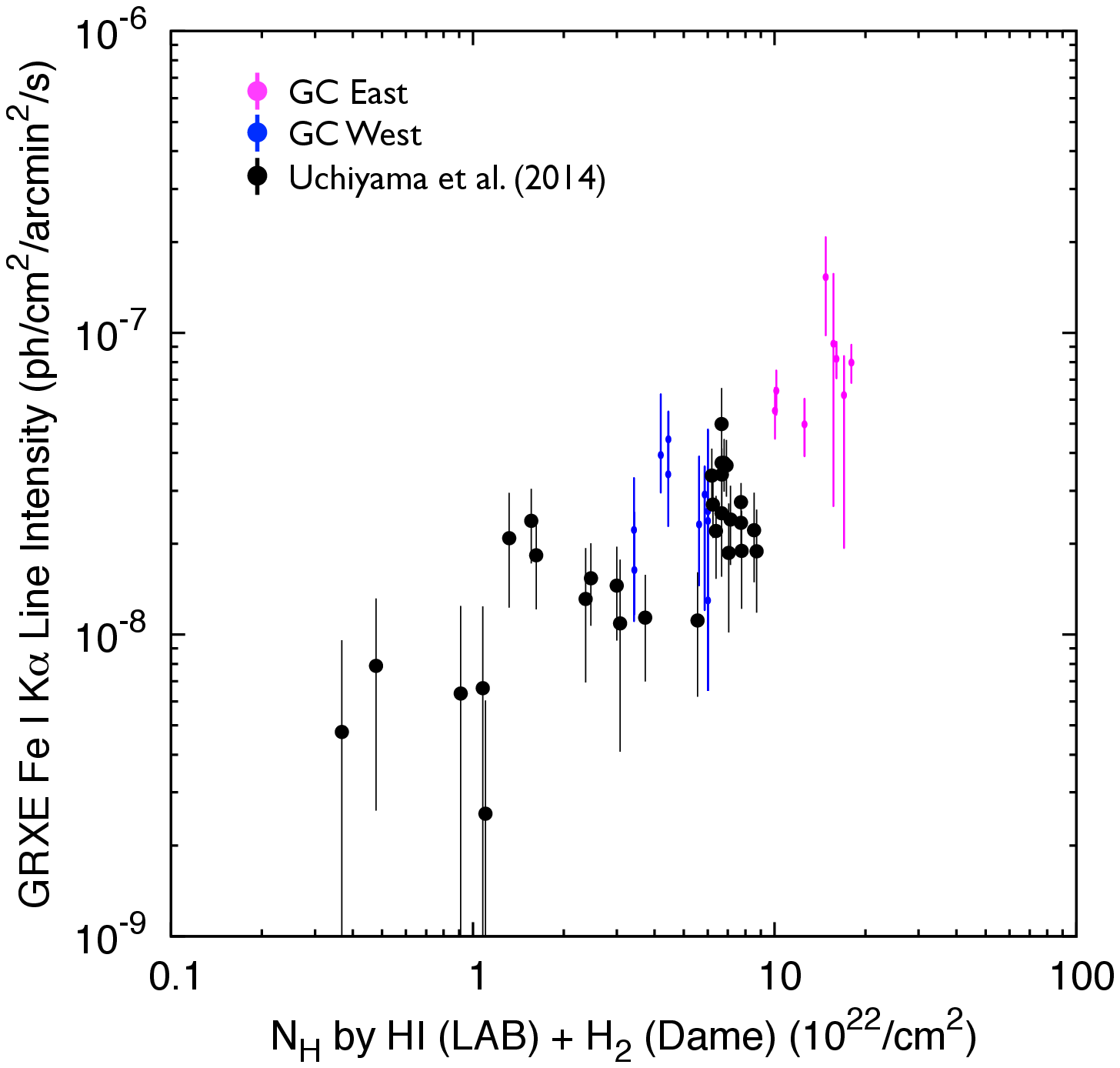}
\caption{Correlation between surface brightness of the 6.4-keV line and the cold ISM column density. 
Magenta and blue points are the data from GC East and West, respectively. 
Black points are the data from $|l|>4^\circ$ that is adopted from Uchiyama et al. (2014). }
\end{figure}

\subsection{Ionizing Particles}


Assuming that the origin of the 6.4-keV line fluorescence from the cold ISM, we now discuss ionizing particles. 
First, we consider the idea of an X-ray reflection nebula, which has already been observed in the Galactic center region. 
For Sgr~A*, observation of the 6.4-keV line and $^{13}$CO in GC East and West regions require a luminosity of $\sim 10^{41} {\rm ergs\ sec^{-1}}$. 
The studies of the Galactic center region ($|l|<1.5^\circ$) imply that Sgr~A* was bright for about 700 yrs at the luminosity of $\sim 10^{39}{\rm ergs\ sec^{-1}}$ (Ryu 2013). 
Thus, the scenario would be possible for the 6.4-keV line also in GC East and West. 
However, maintaining a linear relationship through the entire Galactic plane shown in Figure~7 requires a monotonic increase in the luminosity of Sgr~A* from $10^{41}$ to $10^{43}{\rm ergs\ sec^{-1}}$ for the last $10^4$~yrs, which is not impossible but seems rather artificial. 
The radiation sources would have to be bright binary sources and/or unresolved sources related to the diffuse 6.7-keV line of GRXE. 
We plan to investigate this possibility by comparing the distribution of the 6.4-keV line with the cold ISM and the X-ray sources. 

Dogiel et al. (2011) and Tatischeff et al. (1999) attribute the continuum X-rays and the 6.4-keV line to the impact of low energy cosmic-ray protons (hereafter LECRp) on the cold ISM in the Galactic center region and Galactic plane. 
Assuming the spectrum of the galactic low energy cosmic-ray ($\sim 10\ {\rm MeV}$) is equal to that given by Ip et al. (1985), we find that a LECRp energy density of $\sim 1\times 10^3 {\rm eV\ cm^{-3}}$ is required. 
Thus, LECRp could be ionizing particles near the Galactic center region such as GC East and West. 
However, their energy density would be too high to explain the 6.4-keV line for the entire Galactic plane.

Valinia et al. (2000) propose the interaction of low energy cosmic-ray electrons (hereafter LECRe) with the cold ISM as the mechanism for the GRXE origin including the 6.4-keV line, and they estimate an energy density of $\sim 1 {\rm eV\ cm^{-3}}$. 
We find that the relationship shown in Figure~7 is consistent with this idea. 

In this paper, we present the first results from the Suzaku key project ``Origin of the 6.4-keV line of the Galactic Ridge X-ray Emission''. 
Observation and data analysis of this key project is still ongoing. 
Our next step is to analyze spectra where features such as equivalent widths and absorption edges are crucial to uncover the origin of the 6.4-keV line and its ionizing particles.

\section*{References}
\re Bania et al. 1977, ApJ, 216, 381 
\re Dame, T.~M. et al.  2001, ApJ, 547, 792 
\re Dogiel V. et al. 2011, PASJ, 63, 535 
\re Ebisawa K. et al. 2008, 60, S223 
\re Ip W.-H. et al. 1985, A\&A, 149, 7 
\re Kalberla, P. et al. 2005, A\&A, 440, 775 
\re Koyama K. et al. 2007, PASJ, 59, S245 
\re Maeda Y. 1998, Doctor Thesis, Kyoto Univ. 
\re Mori H. et al. 2005, 57, 245 
\re Piraino S. et al. 2012, A\&A, 542, L27 
\re Ryu S.G. 2013, Doctor Thesis, Kyoto University. 
\re Takeuchi T. et al. 2010, PASJ, 62, 557 
\re Tatischeff V. et al. 1999, ASPC, 171, 226 
\re The Suzaku Technical Description 
\re Uchiyama H. et al. 2011, 63, S903 
\re Uchiyama H. et al. 2014, in prep.  
\re Valinia A. et al. 2000, ApJ, 543, 733 
\re Yamauchi S. et al. 2009, PASJ, 61, S225 
\label{last}

\end{document}